# Influence of the $U_3O_7$ domain structure on cracking during the oxidation of $UO_2$

L.Desgranges, H. Palancher, M. Gamaléri CEA/DEN/DEC Bat 130 C.E. Cadarache 13108 Saint Paul lez Durance FRANCE
J.S. Micha, CNRS,DSM/INAC/SPrAM,C.E. Grenoble, 17 rue des Martyrs, 38054Grenoble cedex 9FRANCE
Virgil Optasanu, Laura Raceanu, Tony Montesin, Nicolas Creton, ICB, UMR 5209 CNRS Av. Alain Savary, 21078 Dijon Cedex FRANCE

*Abstract*
Cracking is observed when a $UO_2$ single crystal is oxidised in air. Previous studies led to the hypothesis thatcracking occurs once a critical depth of $U_3O_7$ oxidised layer is reached. We present some µLaue X-Ray diffraction results, which evidence that the $U_3O_7$ layer, grown by topotaxy on $UO_2$, is made of domains with different crystalline orientations. This observation was used to perform a modelling of oxidation coupling chemical and mechanical parameters, which showed that the domain patterning induces stress localisation. This result is discussed in comparison with stress localisation observed in thin layer deposited on a substrateand used to propose an interpretation of $UO_2$ oxidation and cracking.

*Introduction*

The oxidation of uranium dioxide is a key process involved at least in two important steps of nuclear industry: fuel fabrication and safety of spent fuel dry storage [1]. Oxidation induces cracking and spalling of the initial $UO_2$ material, which might lead to rod cracking in case of defective fuel in dry storage conditions. The morphological evolution of a $UO_2$ pellet during oxidation was studied by Bae [2], who evidenced two stages: first macro-cracking occurs when $U_4O_9$ and $U_3O_7$ are formed, then micro-cracking is associated with the formation of $U_3O_8$. This description is consistent with the one previously proposed by Tempest et al. [3] in which cracks appear in the uniform layer of $U_3O_7$formed on the surface of a $UO_2$ sample. In situ Environmental Scanning Electron Microscope experiments showed that a $UO_2$ single crystal is transformed in a powder during oxidation by the formation of series of macro-cracks [4]. The formation of these macrocracks had been observed as soon as the oxidised layer formed on the initial $UO_2$ sample reached a critical depth of about 0.4µm. This experimental result was used to propose a criterion for the safe handling of oxidized defective fuel rods [5]. This criterion assumes that the nuclear fuel will be safe as long as no macro-crack is formed. A numerical evaluation of this safe duration was also recently proposed by calculating the time needed to form a critical thickness of the oxidised layer [6].

Cracking of thin films has been widely studied both theoretically and experimentallybecause of its relevance for the semi-conductor industry. Although large elastic strains can be tolerated in very thin films, above some critical thickness the film will act to relieve the strains by generating dislocations, micro-twins, surface instabilities and cracks. The formation of cracks in oxidised $UO_2$shares many similarities with cracking of thin film. Macrocracking of $UO_2$ is equivalent to the 2D periodic cracking of thin film witha regularspacing of similar cracks [see for example[7],[8]].In [4] it was shown that the crack which splits the oxidised layer also penetrates in the $UO_2$ substrate, which was observed on thin film deposited on brittle substrate [[9]]. But some discrepancies exist between $UO_2$ oxidation, which proceeds by topotaxy (growth of a new crystalline phase within the substrate by chemical reaction) and thin film growthwhich proceeds by epitaxy (growth of an external phase on a substrate by deposit).First the interface between a film and its substrate is defined by a sharp change of crystalline lattice at a monolayer scale; in the $UO_2$ on the contrary this interface rather corresponds to an interdiffusion layer whose thickness is at least 5 nm [[10]].Second, $UO_2$ oxidation proceeds at temperatures less than 400°C (in dry storage condition) by oxygen diffusion only [[11]], thus the formation of dislocations, which are usually invoked to describe thin film formation, is not likely to exist in $UO_2$ oxidation.

In order to achieve a better description of $UO_2$ oxidation, a new model was developed in which mechanical stresses are coupled to diffusion processes.The first results obtained with a one dimension version of this model [[12]] showed that the oxidation mechanism was better described with an oxidised layer with tensile stress, i.e. a lattice mismatch due to a smaller unit cell volume in the oxidised layer than in the bulk $UO_2$.This model is only valid for isotropic crystalline phase, which is not the case for $U_3O_7$.
To better describe $UO_2$ oxidation a 3D model is needed. In this paper we describe how such a model was built.
Because $UO_2$ oxidation is a complex phenomenon involving several crystalline phases, which change as a function of temperature, oxidation conditions were fixed to get a simple system. The oxidation of a $UO_2$ single crystal in air at 300°C was then chosen, because only $UO_2$ and $U_3O_7$ phases are observed when cracking occurs [4]. Even in this case, oxidation modelling is still a complex problem because $U_3O_7$ phase has a tetragonal crystalline symmetry and several crystalline orientations of $U_3O_7$ on $UO_2$ are possible. In literature the existence of orientation relationships between $UO_2$ and its oxidation layer is reported [[13]],but the topotaxy relationships at the $UO_2$-$U_3O_7$ interface, at which cracking occurs, still needs to be determined.That is why we first studied them experimentally using X-Ray diffraction. Then our mechanical modelling of $UO_2$ oxidation was modified to take them into account. This modified modelling is then used to discuss whether the existence of a critical depth of $U_3O_7$ is needed for cracking to occur.

*characterisation of $UO_2$-$U_3O_7$ interface by X-Ray diffraction*

Several X-Ray diffraction studies of $UO_2$ oxidation are reported in literature [see for example [14],[15]]. But no decisive information were gained on the topotaxy relationships at the $UO_2$-$U_3O_7$ interface, because $U_3O_7$ appears as very broad peaks on the diffraction pattern as long as $UO_2$ can be observed. This can be explained by a continuous change of $U_3O_7$ c/a ratio from 1 to 1.03, and also by the existence of micro-strains induced by the mismatch between $UO_2$ and $U_3O_7$ unit cells. Therefore a new experimental approach is needed to get a more in depth characterisation of $UO_2$-$U_3O_7$ interface.

We used X-ray diffraction with synchrotron radiation in Laue mode on a $UO_2$ single crystal, because this method allows an exploration of the reciprocal space with a better accuracy than conventional X-Ray diffraction.

*Sample preparation*
The 5*5*1mm $UO_2$ single crystal analysed here, was also used for a study of cracking [4] in which it was oxidised in air at 300°C and characterised at different oxidation times. This single crystal has only (111) oriented natural faces, on which characterisation was performed. After 1+1/2 hours of oxidation, a $U_3O_7$ layer was observedby X-Ray diffraction; taking into account the penetration depth of X-Rays, its thickness was estimated around a few tenths of a µm. Its diffraction peaks in (111) direction were characterised by conventional Bragg-Brentano diffraction, and its (111) peak was recorded in a 2D (ω, 2θ) diffraction pattern. The spot corresponding to the (111) $U_3O_7$ diffractionplan is 10 times wider than the $UO_2$ spot in the ω direction. This suggests some disorientation of the $U_3O_7$ crystallites grown on the $UO_2$ single crystal. In order to better characterise this disorientation, the single crystal was brought to ESRF to perform µ-XRD in Laue mode.

*µ-XRD in Laue mode*
µ-XRD has been performed at the BM32 beamline at the ESRF (Grenoble, France). There, a dedicated set-up enables a 2D XRD -mapping with micrometer spatial resolution [16]. Before the experiment, the incoming polychromatic X-ray beam has been carefully characterised: its energy band pass ranges from 5 to 27 keV and its size was 1*2µm² (horizontal and vertical full width at half maximum). During the measurement, the X-ray beam incidence angle has been kept fixed to 40°.
XRD data have been collected with a two-dimensional MAR165 CCD camera at several locations on the oxidised surface. A typical image is shown on Figure 1;it is compared to that collected on a $UO_2$ single crystal previously annealed at 1700°C during 24hours, considered as a strain free reference sample.

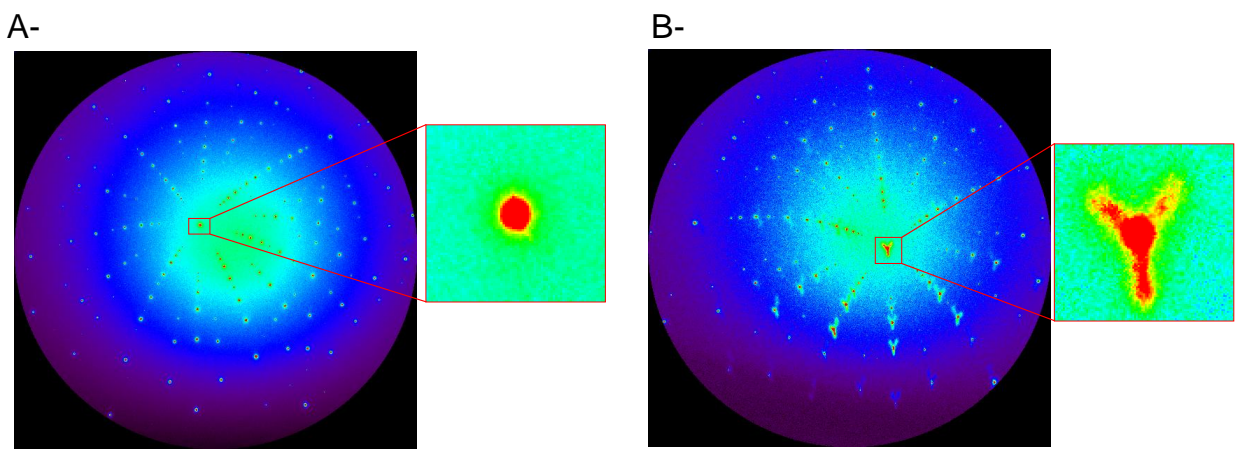

*Figure 1.Comparison of the µ-XRD Laue images collected on a strain free (A-) $UO_2$ and on an oxidised (B-) $UO_2$ single crystals*

The shape of the Bragg spot,initially round on the reference $UO_2$, became elongated along three different directions. These elongations were interpreted considering the crystalline orientations of the $U_3O_7$ crystallites on $UO_2$ substrate.

*Data analysis*

Considering the X-ray penetration depth of X-ray beams into $UO_2$ in the available X-Ray range (5 to 27 keV), both the oxidised layer and the preserved substrate are always probed. The collected image can thus be considered as the sum of two diagrams, the first representative of unstrained $UO_2$ (central round part of the Bragg spot) and the second of the oxidised layer (elongations).

At a first step, using the XMAS software [17], the peak search parameters were set to only take into account the central part of the Bragg peak (i.e. the unstrained one). Based on the obtained peak list, an indexation was performed and the orientation matrix calculated. This indexation confirmed that the $UO_2$ [111] axis was perpendicular to the single crystal surface within several degrees. The position of the (-7,7,7) diffraction spot at the centre of the diffraction pattern in Figure 2 confirms this orientation.

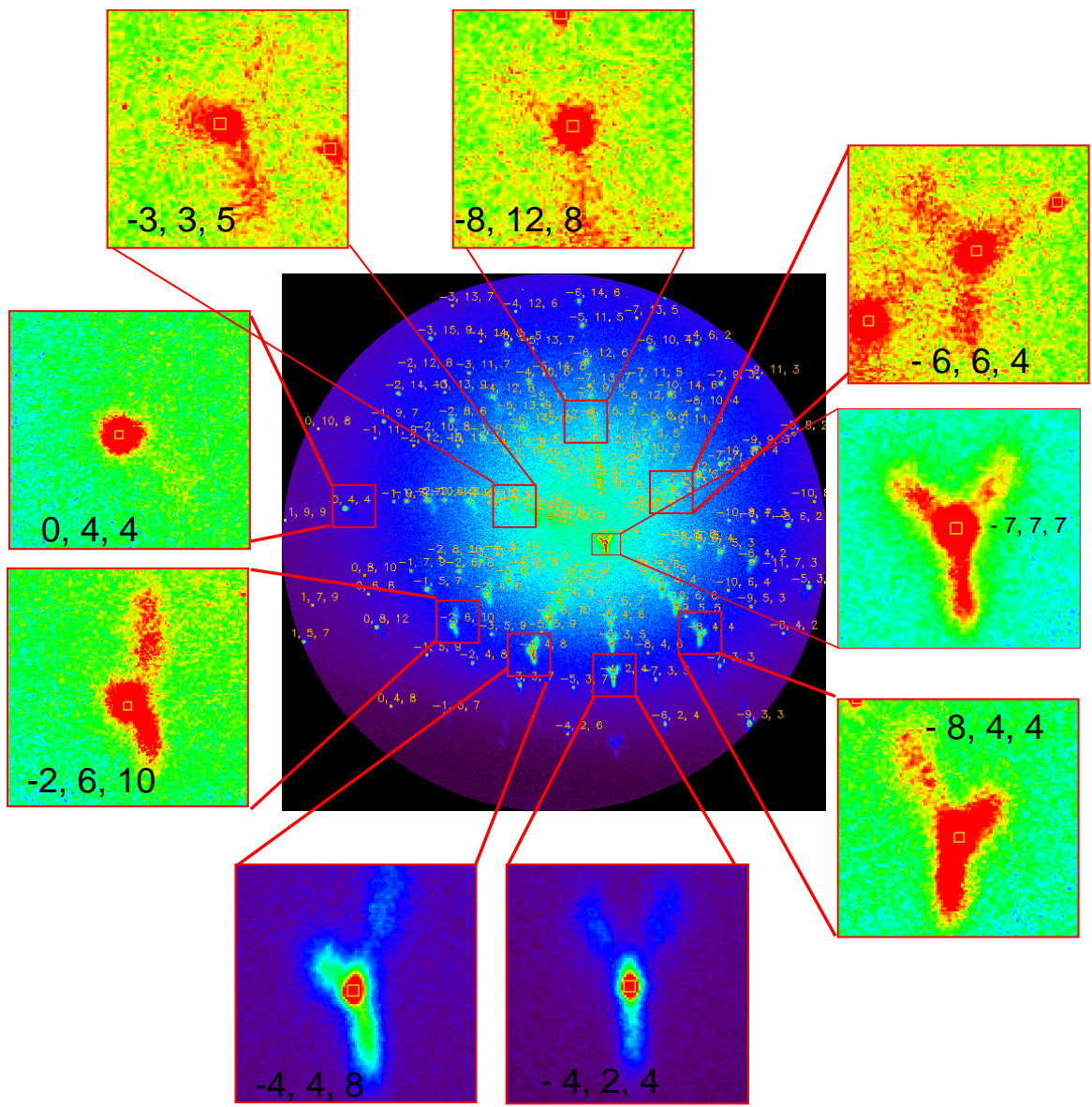

*Figure 2 : Analysis of the image collected on the oxidised $UO_2$ single crystals. Indexation (A-) and simulations of images implying a strained UO2 structure*

At a second step, the elongations were simulated as strained $UO_2$ using the Laue tool software[18]. These simulations allowed determining the directions and intensities of the measured strain.

Each elongation on Figure 2 was successfully attributed to a distorted $UO_2$ lattice. The relationships between $UO_2$ and $U_3O_7$ unit cell used in order to describe the $U_3O_7$ as a distorted $UO_2$ are presented in Figure 3. Both phases share the same crystalline axis. Going for $UO_2$ to $U_3O_7$ leads to a relative increase of 3% along one axis, the elongated axis becomes c axis of $U_3O_7$ structure. Three choices for the $U_3O_7$ c axis are possible, corresponding to the three elongations observed around the $UO_2$ round spot on Figure 2.

The measured distortion is consistent with the $U_3O_7$ crystalline structure, whose c/a ratio is reported to be 1.03 [1]. Moreover $UO_2$ and $U_3O_7$ unit cell axis sharing the same orientation is consistent with the topotactic growth of $U_3O_7$ on $UO_2$. However the intensity of the three elongations shows some fluctuations that can not be explained if the three types of domains were given the same weight. These intensity fluctuations could be interpreted as the $U_3O_7$ domain distribution within the sampling surface of the X-Ray beam, equal to 2 µm². Local variations of the domain microstructure due to crystal surface defects could also lead to intensity fluctuations.

.

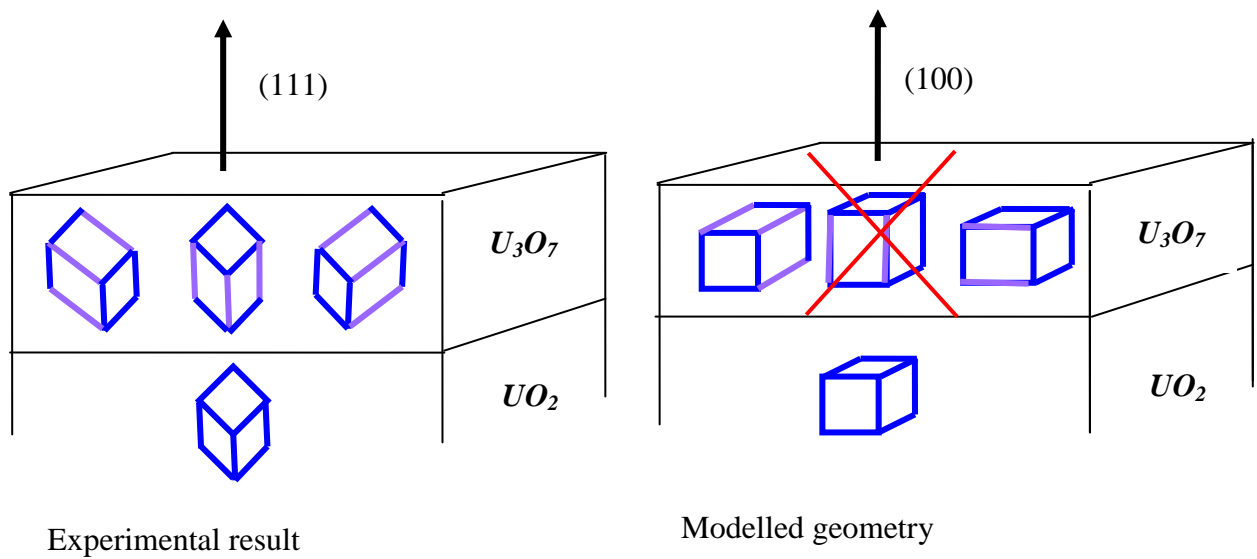

*Figure 3 : Schematic description of the crystalline relationships between $UO_2$ and $U_3O_7$ experimentally observed and modelled.*

### *Modelling $U_3O_7$ layer on $UO_2$ substrate*

In the experimental part it has been shown that $U_3O_7$ is formed with crystallites having different crystalline orientations on the (111) oriented natural face of a $UO_2$ crystal, related to the choice of the c axis when transforming the cubic $UO_2$ unit cell in the tetragonal $U_3O_7$ unit cell. The choice of (111) surface for $UO_2$ implies three different orientations of $U_3O_7$ crystallites. For the modelling purpose, these three different orientations would have led to a complicated geometry. So we decided to describe $UO_2$ oxidation on a (100) surface, because it leads to a simpler geometry.

*hypothesis*

We consider multi-domains of $U_3O_7$ lying on a $UO_2$ substrate. The direction (001) in $UO_2$ is parallel to the direction (010) in $U_3O_7$. Thus, the $c(U_3O_7)$ direction lies in the plane of the interface. We excluded the case in which $c(U_3O_7)$ direction is perpendicular to interface because it corresponds to the highest mismatch between $UO_2$ and $U_3O_7$ (Figure 3). This assumption is consistent with some previous X-Ray diffraction results [15]. The $U_3O_7$ domains have two possible directions for the axis c. The $U_3O_7$ layer is made of alternate domains in which a and c axis contained in the interface permute. The boundaries between these domains are located on (101) planes for energy minimisation. In Figure 4 we show the mesh of our FEM model. The $U_3O_7$ domains are in red and green; the white mesh represents $UO_2$ substrate.

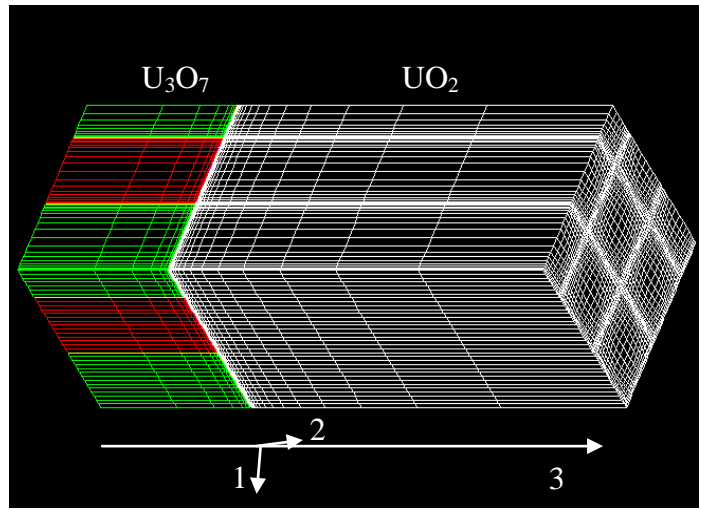

*Figure 4 : An example of mesh. Multi-domain of $U_3O_7$ on $UO_2$ substrate*

This geometry was implemented in the Finite Element Model (FEM) code CASTEM, we used for our calculations. The $U_3O_7$ domains were not modified during the calculations: no mobility of the interfaces was simulated. For a given thickness of $U_3O_7$ layer, the stress distribution and the oxygen diffusion fluxes are successively computed.

The first calculation step computes the stress distribution induced by the connection of different $U_3O_7$ domains and $UO_2$ substrate. The second step computes the oxygen chemical diffusion coupled to the previously calculated mechanical state. The background theory for coupling chemical diffusion to mechanical state was widely exposed in [19], [20], [21], and is based on previous works [22]. The diffusion coefficient we used depends on both the mechanical stress and the stress gradient:

$$D = D_0 \left[ 1 - \frac{M_0 \eta_{ij} c}{RT} \left( \frac{S_{ij}}{c} + \frac{\partial S_{ij}}{\partial c} \right) \right]$$

where $M_0$ is the molar mass of the $UO_2$, $\eta_{ij}$ is the chemical expansion coefficient, $c$ is the oxygen concentration which is dissolved in $UO_2$ lattice, $\sigma$ is the mechanical stress, $T$ is the temperature and $R$ is the universal constant of gases.

Table 1 shows the parameters used in our simulations.

|  | $D_0$ [cm$^3$/s] | $\eta_{ij}^1$ [m$^3$/kg] | T [°C] | E [GPa] | ν | Lattice parameters [Å] | | |
|---|---|---|---|---|---|---|---|---|
|  |  |  |  |  |  | a | b | c |
| Ref. | [23] | [24] |  | [25] |  | [26, 13] | | |
| $UO_2$ | $0.0055 \exp\left(\frac{-26.3}{RT}\right)$ | $-1.248 \cdot 10^{-5}$ | 300 | 200 | 0,32 | 5,47 (cubic) | | |
| $U_3O_7$ |  |  |  |  |  | 5,363 (tetragonal) |  | 5,531 |

*Table 1 : parameters used for computations*

$U_3O_7$ material exists mainly in powder form for which mechanical parameters are difficult to obtain. Thus we assumed that the $U_3O_7$ Young modulus and Poisson's ratio were the same as those of $UO_2$. Chemical expansion coefficient was not considered in $U_3O_7$.

The lateral size of each domain of $U_3O_7$ is 100 nm. Their thickness was fixed at 50 nm, 100 nm, 200 nm, 400 nm successively while the thickness of $UO_2$ is considered as infinite.

*Results*

The stresses calculated in the first computation step is presented in

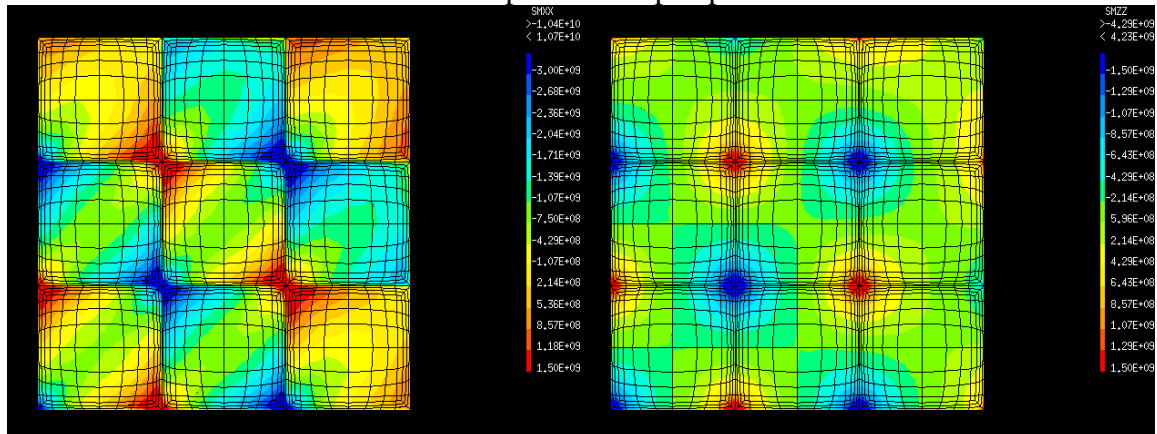

Figure 5 on a section parallel to the $UO_2/U_3O_7$ interface at 1 nm depth inside $UO_2$.

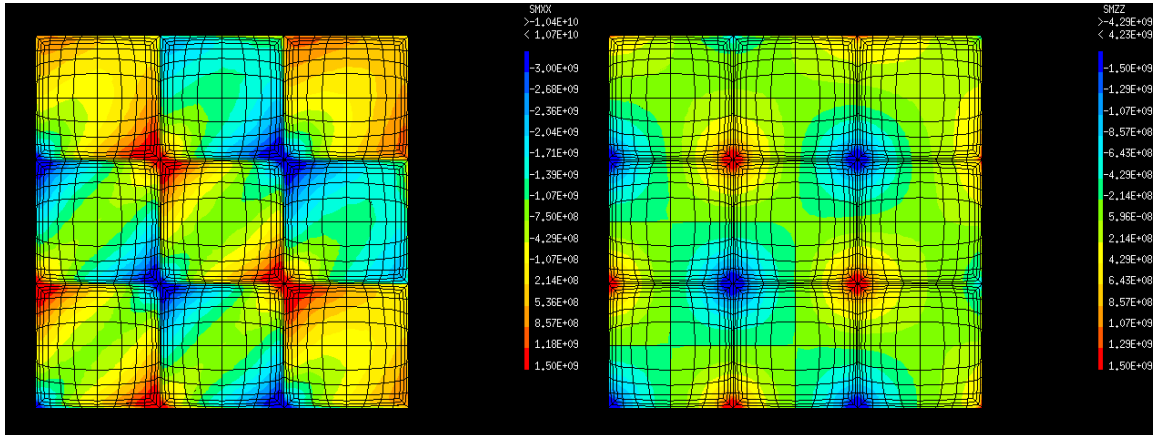

*Figure 5 : $\sigma_{11}$ and $\sigma_{33}$ stresses on a section parallel to the $UO_2/U_3O_7$ interface at 1 nm depth inside $UO_2$.*

The second step computation evidenced that heterogeneities in stress led to different oxygen fluxes in $UO_2$. This produces a transversal oxygen concentration gradient. Moreover, differences in concentration induce differences in chemical expansion, which generates additional mechanical stresses, from -70 MPa to 220 MPa, having a strong feedback on the diffusion. The oxygen concentration distribution is shown in Figure 6 in a transversal section parallel to the $UO_2/U_3O_7$ interface at 10 nm depth in $UO_2$. This oxygen concentration leads to changes of chemical composition in the $UO_{2+x}$ phase with x varying 0.01 to 0.11.

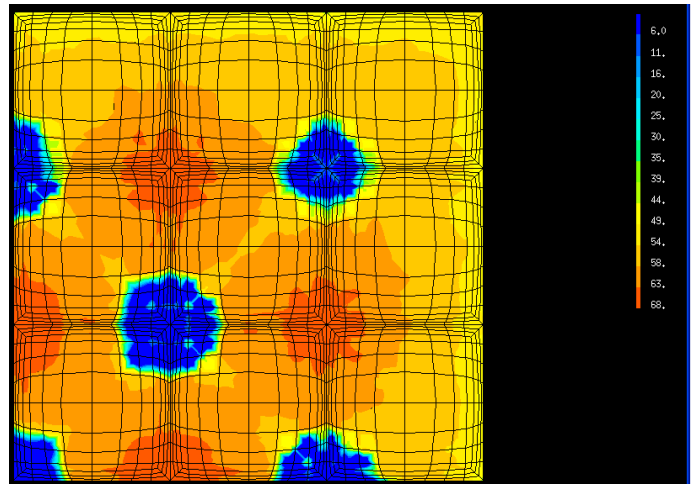

*Figure 6 : Concentration after diffusion in a transversal section $UO_2$-side.*

## *Discussion*

Our experimental results evidenced that the formation of a $U_3O_7$ layer on a $UO_2$ single crystal is associated with the formation of $U_3O_7$ domains having different crystalline orientations related to the choice of $U_3O_7$ c axis. Moreover our model evidenced that the existence of periodical $U_3O_7$ domains induces variations in stresses and oxygen concentration at the $UO_2/U_3O_7$ interface. In the following we discuss why $U_3O_7$ is formed with a domain microstructure, and how this microstructure can justify the existence of a critical depth needed for the formation of cracks in the oxidised layer.

$U_3O_7$ is a metastable phase: in thermodynamic equilibrium conditions without stresses, it is transformed in a mixture of $U_4O_9$ and $U_3O_8$. One reason for the stabilisation of $U_3O_7$ can be the stress state generated by the formation of an oxidised layer on the $UO_2$ substrate.

$U_4O_9$ has a cubic crystalline structure related to the one of $UO_2$; their unit cell parameter differs from 5.47 for $UO_2$ to 5.44 for $U_4O_9$. As $U_4O_9$ is formed by topotaxy on $UO_2$, this difference in cell parameter induces stresses that can be accommodated by an elastic deformation in which $U_4O_9$ would be stretched in the interface plane and compressed perpendicular to it (assuming a Poisson coefficient ν=0.3).

$U_3O_8$ derives from $UO_2$ by a ~ 1/3 expansion along one $UO_2$ (111) axis [11]. In a uniformly oxidised layer, the formation of $U_3O_8$ is likely to induce an expansion in a direction perpendicular to the surface, which minimises the strain state at the interface with the substrate. But the compression stress in $U_4O_9$ makes the structural transformation in $U_3O_8$ more energetically costly.

The minimisation of the mechanical energy of an oxidised layer on a substrate can also be achieved by stress localisation, as in the case of thin layer deposited on a substrate. In this case, Asaro & Tiller [27], and Grinfeld' [28] showed that a sinusoidal deformation of the surface may decrease the mechanical energy of the system. This sinusoidal deformation creates hills and valleys with different mechanical states, stress relieved at the top of the hills and stress increased at the bottom of the valleys. In the case of $UO_2$ oxidation, stress localisation would be achieved by the formation a patterned $U_3O_7$ layer, which minimises the mechanical energy by a reorganisation of crystalline structure rather than by surface deformation.

Thus the formation $U_3O_7$ domains with different crystalline orientations on a $UO_2$ substrate should be understood as a consequence of the minimisation of mechanical energy. Domain formation is a known energy minimisation phenomenon, also for example in magnetism [29].

In the case of thin layer deposited on a substrate, stress localisation leads to periodical cracking of the sample at a critical depth because the stress level increases with increasing depth of the layer. In the case of $UO_2$ oxidation we performed calculations with several thickness of $U_3O_7$ layer in order to check the stress intensity as a function of $U_3O_7$ thickness. In .Figure 7 $\sigma_{11}$ stress is shown as a function of the distance to the $UO_2/U_3O_7$ interface. The maximal value of the stress increases as a function of the $U_3O_7$ thickness.

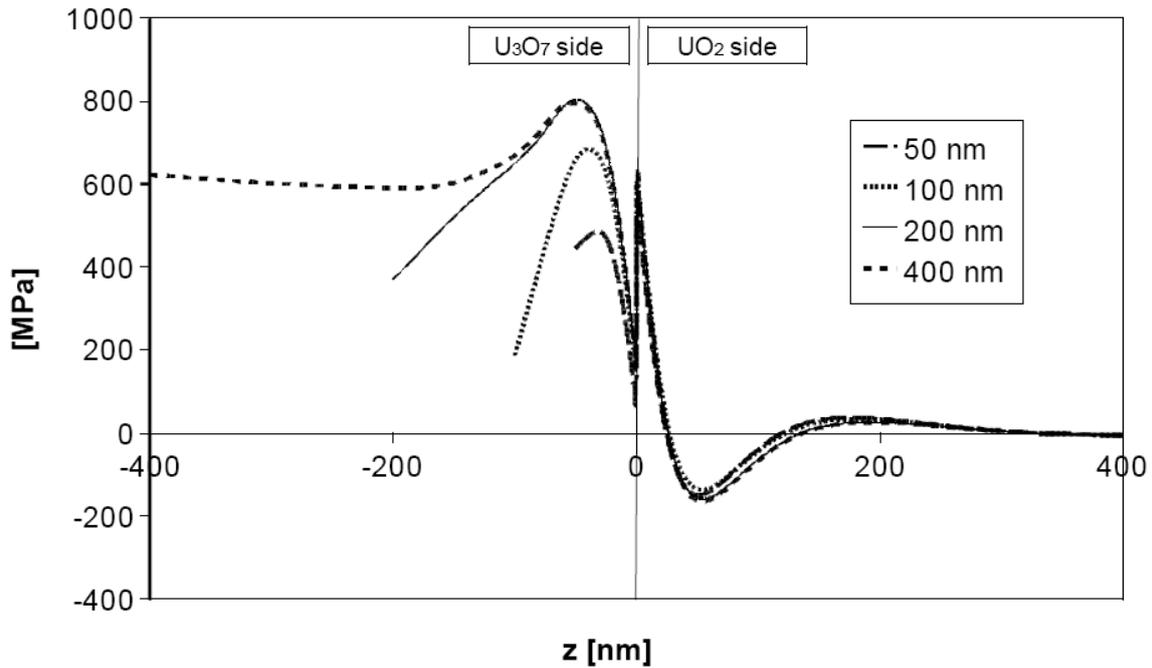

*Figure 7 $\sigma_{11}$ stress for several thickness of $U_3O_7$ layer.*

The increase of traction stress with $U_3O_7$ layer depth justifies that this layer cracks when its thickness reaches a critical value. No exact value of the critical thickness can be derived from our calculation, because the numerical values we used are, for some them, crude approximations. The formation of cracks changes the mechanical state of the layer and would allow the formation of $U_3O_8$, as it is observed experimentally. Considering a martinsitic type formation of $U_3O_8$ on $UO_2$ as proposed by Allen [13], these $U_3O_8$ crystallites should have a size smaller or equal to the one of the $U_3O_7$ grain, they were formed from.

The mechanism described here is consistent with all reported results on $UO_2$ at temperature less than 400°C. It could also be applicable to a wider class of materials which are submitted to cracking during a chemical reaction, a phenomenon sometimes named chemical fragmentation [30].

*Conclusion :*

$UO_2$ oxidation in air at temperature less that 400°C shares some common features with thin film deposited on substrate. The topotactic formation of $U_3O_7$ oxidised layer can indeed be compared to the epitaxial formation of a thin layer. In both case the mismatch in unit cell parameter between the layer and the substrate drives the mechanical evolution of the system. Because uranium oxide does not allow surface deformation, stress localisation, which minimises mechanical energy, is achieved because of the formation of domains of the metastable $U_3O_7$ phase. This original mechanism can explain the existence of a critical depth for crack formation and could be applied to materials subject to "chemical fragmentation".
10

**Acknowledgements:** The authors would like to acknowledge the financial support from Electricité de France.